\documentclass[11pt]{article}
\usepackage{graphicx}
\usepackage{amsmath}
\usepackage{amssymb}

\begin{document}

\title{Propagators of the Dirac fermions in the de Sitter expanding universe}

\author{{Ion I. Cot\u aescu\footnote{E-mail: i.cotaescu@e-uvt.ro}}\\
{\small \it West University of Timi\c{s}oara,}\\
{\small \it V.  P\^{a}rvan Ave.  4, RO-300223 Timi\c{s}oara, Romania}}

\maketitle

\begin{abstract}
The propagators of the Dirac fermions are studied in the configuration representation on the expanding portion of the $(1+3)$-dimensional de Sitter spacetime considering a fixed vacuum of Bunch-Davies type. In this representation the method of Koksma and Prokopec [J. F. Koksma and T. Prokopec, Class. Quant. Grav. {\bf  26}, 125003 (2009)] is applied recovering thus the form of the propagators in the massive case but obtaining a new result  for the left-handed massless fermions (neutrinos). 
\end{abstract}


\maketitle
\newpage
\section{Introduction}

Important pieces of the quantum field theory on curved spacetimes are the two-point functions that can be calculated either as propagators by using  mode expansions or by looking for new hypotheses complying with the general relativistic covariance as, for example, that of the maximal symmetry of the two-point functions on the hyparbolic spacetimes, i. e. de Sitter (dS) and anti-de Sitter ones \cite{SW}.  

The propagator of the Dirac fermions on the dS spacetime in configuration representation was derived first by Candelas and Reine which  integrated the Green equation of this field \cite{CR}. The same propagator was calculated later as mode sum by Koskma and Prokopec in a more general context of spacetimes of arbitrary dimensions approaching to the dS one \cite{KP}. On the other hand, we developed the dS QED in Coulomb gauge \cite{CQED} where we know the Dirac quantum modes in different bases \cite{CD1,CD2,CD3} and we need to derive the Feynman propagators for calculating physical effects. Obviously, its expression in configuration representation is included in the general result of Ref. \cite{KP} but it must be extracted in the particular case of the $(1+3)$-dimensional dS spacetime and a fixed vacuum. 

This is a good opportunity for reviewing the entire procedure of Ref. \cite{KP} of deriving  the propagators of the Dirac field in configuration representation  on the expanding portion of the dS spacetime. We assume that  the covariant Dirac field is quantized canonically \cite{CdSquant} and the vacuum is fixed being of the Bunch-Davies type \cite{BuD,BD} as in our dS QED \cite{CQED}.  Our goal is to present all the calculation details discussing the specific properties of the quantities under consideration.    For thechnical rerasons, we rewrite the theory of the free Dirac field in terms of modified Bessel functions \cite{NIST,GR} instead of the Hankel ones used in our previous papers. Thus we recover the results of Ref. \cite{KP} for the massive fermions in the particular case of the $(1+3)$-dimensional dS manifold but we obtain a different result for the  left-handed massless fermions.  This is because the left-handed fermions (neutrinos) can be defined only in $(1+3)$ dimensions without generalizations to higher dimensions. 

This paper is a short technical revew consiting of five sections. In the next one we review the fundamental solutions of the Dirac equation minimally coupled to the dS gravity in the massive and massless cases. We point out that these form complete systems of orthonormalized spinors allowing one to write down the mode expansion of the Dirac field. The third section is devoted to the anti-commutator matrix-functions and  propagators defined as mode sums which can be calculated applying the method Ref. \cite{KP}. In the next one we discuss the form of the fermion propagators in the configuration representation in the massive and massless cases as resulted from the technical ingredients presented in the Appendices A and B. Here we observe that our propagator of the  left-handed massless fermions is different from that of Ref. \cite{KP}.  Finally we present some concluding remarks.

\section{Fundamental spinor solutions}

Let us first revisit some basics properties of the fundamental solutions of the  Dirac equation minimaly coupled to the gravity of the $(1+3)$-dimensional de Sitter expanding universe. In what follows we consider the normalized solutions of positive and negative frequencies of the spin basis \cite{CD3} since those of the helicity basis \cite{CD1} are not defined in rest frames. 

\subsection{Dirac spinors in conformal charts}

We denote by $M$  the de Sitter expanding universe of radius $\frac{1}{\omega}$ where the notation $\omega$ stands for its Hubble constant. We choose the moving chart $\{x\}=\{t,\vec{x}\}$ of the {\em conformal} time, $t\in (-\infty,0]$, Cartesian coordinates and
the line element
\begin{equation}
ds^{2} =\frac{1}{(\omega t)^2}\,\left(dt^{2}- d\vec{x}\cdot
d\vec{x}\right)\,,
\end{equation}
which covers the expanding part of the de Sitter manifold. In
addition, we use the non-holonomic frames defined by the tetrad
fields which have only diagonal components,
\begin{equation}\label{tt}
e^{0}_{0}=-\omega t\,, \quad e^{i}_{j}=-\delta^{i}_{j}\,\omega t \,,\quad \hat
e^{0}_{0}=-\frac{1}{\omega t}\,, \quad \hat e^{i}_{j}=-\delta^{i}_{j}\,
\frac{1}{\omega t}\,.
\end{equation}

In this tetrad-gauge, the massive Dirac field $\psi$ of mass $m$ and its Dirac adjoint $\bar{\psi}=\psi^+\gamma^0$ satisfy the field equations $(D_x-m) \psi (x)=0$ and, respectively, $\bar{\psi}(x)(\bar{D}_x-m)=0$ given by the Dirac operator 
\begin{equation}\label{ED}
D_x=-i\omega t\left(\gamma^0\partial_{t}+\gamma^i\partial_i\right)
+\frac{3i\omega}{2}\gamma^{0}\,,
\end{equation}
and its adjoint 
\begin{equation}\label{ED1}
\bar{D}_x=\left(\gamma^0\stackrel{\leftarrow}\partial_{t}+\gamma^i \stackrel{\leftarrow}{\partial}_i\right)i\omega t
-\frac{3i\omega}{2}\gamma^{0}\,,
\end{equation}
whose derivatives act to the left. Note that in the frames we work  the Dirac operator has the property $D_{-x}=D_x$ which can be used in applications.
%%%%

The quantum Dirac field may be expanded in terms of fundamental spinors of positive and negative frequencies in different representations. Here we consider the momentum representation where the plane wave solutions  $U_{\vec{p},\sigma}$  and  $V_{\vec{p},\sigma}$ depend on the momentum $\vec{p}$ and arbitrary polarization $\sigma$. These spinors  form an orthonormal basis satisfying the orthogonality relations
\begin{eqnarray}
&&\langle U_{\vec{p},\sigma}, U_{{\vec{p}\,}',\sigma'}\rangle=
\langle V_{\vec{p},\sigma}, V_{{\vec{p}\,}',\sigma'}\rangle =
\delta_{\sigma\sigma^{\prime}}\delta^{3}(\vec{p}-\vec{p}\,^{\prime})\label{U}\\
&&\langle U_{\vec{p},\sigma}, V_{{\vec{p}\,}',\sigma'}\rangle=
\langle V_{\vec{p},\sigma}, U_{{\vec{p}\,}',\sigma'}\rangle =0\,, \label{V}
\end{eqnarray}
with respect to the relativistic scalar product \cite{CD1}
\begin{equation}
\langle \psi, \psi'\rangle=\int d^{3}x
\sqrt{|g|}\,e^0_0\,\bar{\psi}(x)\gamma^{0}\psi(x)=\int d^{3}x
(-\omega t)^{-3}\bar{\psi}(x)\gamma^{0}\psi(x)\,, 
\end{equation}
and the completeness condition  \cite{CD1}
\begin{equation}\label{complet}
\int d^{3}p
\sum_{\sigma}\left[U_{\vec{p},\,\sigma}(t,\vec{x}\,)U^{+}_{\vec{p},\sigma}(t,\vec{x}\,^{\prime}\,)+
V_{\vec{p},\sigma}(t,\vec{x}\,)V^{+}_{\vec{p},\sigma}(t,\vec{x}\,^{\prime}\,)\right]=(-\omega t)^3\delta^{3}(\vec{x}-\vec{x}\,^{\prime})\,,
\end{equation}
In this representation the Dirac field may be expanded as 
\begin{eqnarray}
\psi(t,\vec{x}\,) & = &
\psi^{(+)}(t,\vec{x}\,)+\psi^{(-)}(t,\vec{x}\,)\nonumber\\
& = & \int d^{3}p
\sum_{\sigma}[U_{\vec{p},\sigma}(x)a(\vec{p},\sigma)
+V_{\vec{p},\sigma}(x)b^{+}(\vec{p},\sigma)]~,\label{p3}
\end{eqnarray}
assuming that the particle $(a,a^{\dagger})$ and antiparticle ($b,b^{\dagger})$
operators  satisfy the canonical commutation relations \cite{CD1,CdSquant}, 
\begin{equation}
\{a(\vec{p},\sigma),a^{+}(\vec{p}\,\,^{\prime},\sigma^{\prime})\}=
\{b(\vec{p},\sigma),b^{+}(\vec{p}\,\,^{\prime},\sigma^{\prime})\}=\delta_{\sigma\sigma^{\prime}}
\delta^{3}(\vec{p}-\vec{p}\,^{\prime})\,.
\end{equation}
Thus we obtain a good quantum theory where the one-particle operators conserved via Noether theorem become just the generators of the corresponding isometries \cite{CdSquant}. 

The plane wave solutions can be derived as in Refs. \cite{CD1,CD3} by solving the Dirac equation in the standard representation of the Dirac matrices (with diagonal $\gamma^0$). It is convenient to express here these solutions in terms of modified Bessel functions  $K_{\nu}$ instead of Hankel functions as in Refs.  \cite{CD3,CdSquant}. Thus, by choosing new suitable phase factors we may write
\begin{eqnarray}
U_{\vec{p},\sigma}(t,\vec{x}\,)&=& \sqrt{\frac{p}{\pi\omega}}\,(\omega t)^2\left(
\begin{array}{c}
K_{\nu_{-}}(ip t) \,
\xi_{\sigma}\\
K_{\nu_{+}}(ip t) \,
 \frac{\vec{p}\cdot\vec{\sigma}}{p}\,\xi_{\sigma}
\end{array}\right)
\frac{e^{i\vec{p}\cdot\vec{x}}}{(2\pi)^{\frac{3}{2}}}\label{Ups}\\
V_{\vec{p},\sigma}(t,\vec{x}\,)&=&-\sqrt{\frac{p}{\pi\omega}}\, (\omega t)^2 \left(
\begin{array}{c}
K_{\nu_{-}}(-i p t)\,
\frac{\vec{p}\cdot\vec{\sigma}}{p}\,\eta_{\sigma}\\
K_{\nu_{+}}(-i p t) \,\eta_{\sigma}
\end{array}\right)
\frac{e^{-i\vec{p}\cdot\vec{x}}}{(2\pi)^{\frac{3}{2}}}\,,\label{Vps}
\end{eqnarray}
where $p=|\vec{p}|$ and $\nu_{\pm}=\frac{1}{2}\pm i\mu$, with $\mu=\frac{m}{\omega}$.  The  the Pauli spinors $\xi_{\sigma}$ and $\eta_{\sigma}= i\sigma_2 (\xi_{\sigma})^{*}$ must be correctly normalized,  $\xi^+_{\sigma}\xi_{\sigma'}=\eta^+_{\sigma}\eta_{\sigma'}=\delta_{\sigma\sigma'}$,  satisfying the completeness condition \cite{BDR}
\begin{equation}\label{Pcom}
\sum_{\sigma}\xi_{\sigma}\xi_{\sigma}^+=\sum_{\sigma}\eta_{\sigma}\eta_{\sigma}^+={\bf 1}_{2\times 2}\,.
\end{equation}
In Ref. \cite{CD1} we used the Pauli spinors of the {\em helicity} basis in which the direction of the spin projection is just that of the momentum $\vec{p}$. However, we can project the spin on an arbitrary direction, independent on $\vec{p}$, as in the case of the
{\em spin} basis  where the spin is projected on the third axis of the rest frame such that  $\xi_{\frac{1}{2}}=(1,0)^T$ and $\xi_{-\frac{1}{2}}=(0,1)^T$ for particles and $\eta_{\frac{1}{2}}=(0,-1)^T$ and $\eta_{-\frac{1}{2}}=(1,0)^T$ for
antiparticles \cite{CD3}.

In the case  $m=0$ (when $\mu=0$) it is convenient to consider the chiral representation of the Dirac matrices (with diagonal $\gamma^5$) and the momentum-helicity basis in the chart $\{t,\vec{x}\}$. Then the fundamental solutions, $U_{\vec{p}^0,\lambda}$ and $V_{\vec{p}^0,\lambda}$, of the left-handed massless Dirac field can be written as  \cite{CD1},
\begin{eqnarray}
U^0_{\vec{p},\lambda}(t,\vec{x})&=&
\lim_{\mu\to 0} \frac{1-\gamma^5}{2} U_{\vec{p},\lambda}(t,\vec{x})
\nonumber\\
&=&\left(\frac{-\omega t}{2\pi}\right)^{3/2}
\left(
\begin{array}{c}
(\frac{1}{2}-\lambda)\tilde\xi_{\lambda}(\vec{p})\\
0
\end{array}\right)
\,e^{-ipt+i\vec{p}\cdot\vec{x}}\,, \label{n1}\\
V^0_{\vec{p},\lambda}(t,\vec{x})&=&
\lim_{\mu\to 0} \frac{1-\gamma^5}{2}V_{\vec{p},\lambda}(t,\vec{x})
\nonumber\\
&=&\left(\frac{-\omega t}{2\pi}\right)^{3/2}
\left(
\begin{array}{c}
(\frac{1}{2}+\lambda)\tilde\eta_{\lambda}(\vec{p})\\
0
\end{array}\right)
\,e^{ipt-i\vec{p}\cdot\vec{x}}\,.  \label{n2}
\end{eqnarray}
We observe that the only non-vanishing components are either of positive frequency and $\lambda=-1/2$ or of negative frequency and $\lambda=1/2$, as in Minkowski spacetime. This is because the massless Dirac equation is conformal covariant such that the dS spinors are just  the Minkowski ones multiplied with the conformal factor $(-\omega t)^{\frac{3}{2}}$.  

\subsection{Orbital and spin terms}

The form of the spinors (\ref{Ups}) and (\ref{Vps}) suggests us to introduce the auxiliary $4\times 4$  matrix-functions
\begin{equation}
W_{\pm}(x)=\left(
\begin{array}{cc}
K_{\nu_{\pm}}(i x)&0\\
0&K_{\nu_{\mp}}(i x)
\end{array}\right)\,,\quad \forall x\in {\Bbb R}\,, 
\end{equation}
which have the obvious properties
$\bar{W}_{\pm}(x)=W_{\pm}(x)^*=\gamma^5 W_{\pm}(-x)\gamma^5=W_{\mp}(-x)$ and satisfy 
\begin{equation}
Tr\left[W_{\pm}(x)W_{\mp}(-x)\right]=\frac{2\pi}{ x}\,,
\end{equation}
as it results from Eq. (\ref{H3}) and observing that ${\rm Tr}(\pi_{\pm})=2$.
With their help we can write the fundamental spinors in a simpler form as
\begin{eqnarray}
U_{\vec{p},\sigma}(t,\vec{x}\,)&=& \sqrt{\frac{p}{\pi\omega}}\,\frac{e^{i\vec{p}\cdot\vec{x}}}{(2\pi)^{\frac{3}{2}}}\,(\omega t)^2\, W_-(pt) \gamma(\vec{p}) u_{\sigma}
\label{Ups1}\\
V_{\vec{p},\sigma}(t,\vec{x}\,)&=&\sqrt{\frac{p}{\pi\omega}}\, 
\frac{e^{-i\vec{p}\cdot\vec{x}}}{(2\pi)^{\frac{3}{2}}}\,(\omega t)^2\,W_-(-pt) \gamma(\vec{p})v_{\sigma}\,,\label{Vps1}
\end{eqnarray}
depending on the nilpotent matrix
\begin{equation}
\gamma(\vec{p})=\frac{\gamma^0 p-\vec{\gamma}\cdot\vec{p}}{p}\,, 
\end{equation}
and the 4-dimensional spinors 
\begin{equation}
u_{\sigma}=\left(
\begin{array}{c}
\xi_{\sigma}\\
0
\end{array}\right)\quad
v_{\sigma}=\left(
\begin{array}{c}
0\\
\eta_{\sigma}
\end{array}\right)
\end{equation}
that allow us to define the usual projector matrices
\begin{equation}
\pi_+=\sum_{\sigma}u_{\sigma}\bar{u}_{\sigma}=\frac{1+\gamma^0}{2}\,,\quad \pi_-=\sum_{\sigma}v_{\sigma}\bar{v}_{\sigma}=\frac{1-\gamma^0}{2}\,,
\end{equation}
that form a complete system since $\pi_+\pi_-=0$ and $\pi_++\pi_-=1$. All these auxiliary quantities will help us to perform easily the further calculations either by using the form 
\begin{equation}
W_{\pm}(x)=\pi_+ K_{\nu_{\pm}}(ix)+\pi_-K_{\nu_{\mp}}(ix)\,,
\end{equation}
and simple rules as 
$\gamma(\vec{p})^2=0\,, \gamma(\vec{p})\gamma(-\vec{p})=2\gamma(\vec{p})\gamma^0\,, \gamma(\vec{p})\pi_{\pm}\gamma(\vec{p})=\pm\gamma(\vec{p})$, etc., or resorting to algebraic codes on computer.   

In this formalism we can separate the spin part encapsulated in the terms of the form $\gamma(\vec{p})u_{\sigma}$ and $\gamma(\vec{p})v_{\sigma}$ remaining with the orbital parts that can be concentrated in the new quantities
\begin{equation}\label{w}
w_{\vec{p}}^{\pm}(t,\vec{x})=\sqrt{\frac{p}{\pi\omega}}\,\frac{e^{i\vec{p}\cdot\vec{x}}}{(2\pi)^{\frac{3}{2}}}\,(\omega t)^2\, W_{\pm}(pt)\,,
\end{equation}
which allow us to write the fundamental spinors simply as
\begin{eqnarray}
U_{\vec{p},\sigma}(t,\vec{x}\,)&=& w^-_{\vec{p}}(t,\vec{x}) \gamma(\vec{p}) u_{\sigma}\,,\\
V_{\vec{p},\sigma}(t,\vec{x}\,)&=& w^-_{\vec{p}}(-t,-\vec{x}) \gamma(\vec{p}) v_{\sigma}\,.
\end{eqnarray}
The new matrix-functions (\ref{w}) are reducible and satisfy the remarkable identities
\begin{equation}
\frac{1}{-\omega t}(D_x\pm m)\,w^{\pm}_{\vec{p}}(t,\vec{x})= w^{\mp}_{\vec{p}}(t,\vec{x})p \gamma(\vec{p})\label{Idw}
\end{equation}
and the second order equations
\begin{equation}
(D_x\mp m)\frac{1}{-\omega t}(D_x\pm m)\,w^{\pm}_{\vec{p}}(t,\vec{x})=0\,,
\end{equation}
helping us to recover easily some important results of Ref. \cite{KP}. 

\section{Anti-commutator and Green matrix-functions}

Let us consider the partial anti-commutators matrix-functions of positive and negative frequencies \cite{CD1},
\begin{equation}
{S}^{(\pm)}(t,t^{\prime},\vec{x}-\vec{x}\,^{\prime}\,)=-i\{\psi^{(\pm)}(t,\vec{x})\,,\bar{\psi}
^{(\pm)}(t^{\prime},\vec{x}\,^{\prime}\,)\}\,,\label{Spm}
\end{equation}
which satisfy the Dirac equation in both sets of variables,
\begin{equation}
(D_x - m){S}^{(\pm)}(t,t^{\prime},\vec{x}-\vec{x}\,^{\prime}\,)={S}^{(\pm)}(t,t^{\prime},\vec{x}-\vec{x}\,^{\prime}\,)(\bar{D}_{x'}-m)=0\,.
\end{equation}
The total anti-commutator matrix-function \cite{CD1}
\begin{equation}
{S}(t,t^{\prime},\vec{x}-\vec{x}\,^{\prime}\,)=-i\{\psi(t,\vec{x}\,),\bar{\psi}(t',\vec{x}\,^{\prime}\,)\}={S}^{(+)}(t,t^{\prime},\vec{x}-\vec{x}\,^{\prime}\,)+{S}^{(-)}(t,t^{\prime},\vec{x}-\vec{x}\,^{\prime}\,)\label{Stot}
\end{equation} 
has similar properties and, in addition, satisfy the equal-time condition
\begin{equation}
{S}(t,t,\vec{x}-\vec{x}\,^{\prime}\,)=-i\gamma^0( -\omega t)^3\delta^{3}(\vec{x}-\vec{x}\,^{\prime})
\end{equation}
resulted from Eq. (\ref{complet}). 

In the quantum theory of fields it is important to study  the Green functions
related to the partial or total anti-commutator matrix-functions. We introduce the retarded (R) and advanced (A) Green functions \cite{BDR},
\begin{eqnarray}
S_R(t,t',\vec{x}-\vec{x}\,^{\prime}\,)&=&\theta(t-t')S(t,t',\vec{x}-\vec{x}\,^{\prime}\,)\label{SR}\\
S_A(t,t',\vec{x}-\vec{x}\,^{\prime}\,)&=&-\theta(t'-t)S(t,t',\vec{x}-\vec{x}\,^{\prime}\,)\label{SA}
\end{eqnarray}
while the Feynman propagator is defined in usual manner as \cite{BDR},
\begin{eqnarray}
&&S_{F}(t,t^{\prime},\vec{x}-\vec{x}\,^{\prime})=
-i\langle0|T[\psi(x)\bar{\psi}(x^{\prime})]|0\rangle\nonumber\\
 &&~~~~~~=  \theta(t-t^{\prime})S^{(+)}(t,t^{\prime},\vec{x}-\vec{x}\,^{\prime}\,)-\theta(t^{\prime}-t)
S^{(-)}(t,t^{\prime},\vec{x}-\vec{x}\,^{\prime}).\label{SF}
\end{eqnarray}
From the above definitions we find that these Green functions satisfy the Green equation that in the conformal chart has the form \cite{CD1},
\begin{eqnarray}
(D_x-m)S_{F/R/A}(t,t^{\prime},\vec{x}-\vec{x}\,^{\prime})&=&S_{F/R/A}(t,t^{\prime},\vec{x}-\vec{x}\,^{\prime})(\bar{D}_{x'}-m)\nonumber\\
&=&(-\omega t)^3 \delta^{4}(x-x\,^{\prime})\,. \label{p8}
\end{eqnarray}
However, the Green equation has an infinite set of solutions corresponding to various initial conditions. Here we are interested to study only the Green functions $S_R$, $S_A$ and $S_F$ which will be called propagators in what follows.  

The anti-commutator matrix-functions can be calculated with the help of  the fundamental spinors   (\ref{U}) and (\ref{V}) obtaining similar expressions,
\begin{eqnarray}
&&i{S}^{(+)}(t,t^{\prime},\vec{x}-\vec{x}\,^{\prime}\,)=\sum_{\sigma}\int d^3p\, U_{\vec{p},\sigma}(t,\vec{x}\,)\bar{U}_{\vec{p},\sigma}(t\,^{\prime},\vec{x}\,^{\prime}\,) \nonumber\\
&&~~~~~~~~= \frac{\omega^3}{8\pi^4} (tt')^2\int d^3p\,p\, e^{i\vec{p}\cdot(\vec{x}-\vec{x}')}W_{-}(pt)\gamma(\vec{p})W_{+}(-pt')\,,\label{Splus}\\
&&i{S}^{(-)}(t,t^{\prime},\vec{x}-\vec{x}\,^{\prime}\,)=\sum_{\sigma}\int d^3p\, V_{\vec{p},\sigma}(t,\vec{x}\,)\bar{V}_{\vec{p},\sigma}(t\,^{\prime},\vec{x}\,^{\prime}\,) \nonumber\\
&&~~~~~~~~= \frac{\omega^3}{8\pi^4} (tt')^2\int d^3p\,p\, e^{i\vec{p}\cdot(\vec{x}-\vec{x}')}W_-(-pt)\gamma(-\vec{p})W_+(pt')\,, \label{Smin}
\end{eqnarray} 
after changing $\vec{p}\to-\vec{p}$ in the last integral. Furthermore, we procede as in Ref. \cite{KP} exploiting  Eq. (\ref{Idw}) which allow one to write
\begin{equation}\label{SSig}
{S}^{(\pm)}(t,t^{\prime},\vec{x}-\vec{x}\,^{\prime}\,)=\frac{1}{-\omega t}(D_x+m){\Sigma}^{(\pm)}(t,t^{\prime},\vec{x}-\vec{x}\,^{\prime}\,)\,,
\end{equation}
where the new simpler matrix-functions
\begin{eqnarray}
&&i{\Sigma}^{(+)}(t,t^{\prime},\vec{x}-\vec{x}\,^{\prime}\,) = \frac{\omega^3}{8\pi^4} (tt')^2\int d^3p\, e^{i\vec{p}\cdot(\vec{x}-\vec{x}')}W_+(pt)W_+(-pt')\,, \label{Splus1}\\
&&i{\Sigma}^{(-)}(t,t^{\prime},\vec{x}-\vec{x}\,^{\prime}\,) = - \frac{\omega^3}{8\pi^4} (tt')^2\int d^3p\, e^{i\vec{p}\cdot(\vec{x}-\vec{x}')}W_+(-pt)W_+(pt')\,,\label{Smin1}
\end{eqnarray} 
have the remarkable property
\begin{equation}\label{SiSi}
{\Sigma}^{(-)}(t,t^{\prime},\vec{x}-\vec{x}\,^{\prime}\,)=-{\Sigma}^{(+)}(t^{\prime},t,\vec{x}-\vec{x}\,^{\prime}\,)\,.
\end{equation}
Note that a similar representation can be written as 
\begin{equation}\label{SSig1}
{S}^{(\pm)}(t,t^{\prime},\vec{x}-\vec{x}\,^{\prime}\,)={\bar\Sigma}^{(\pm)}(t,t^{\prime},\vec{x}-\vec{x}\,^{\prime}\,)(\bar{D}_x+m)\frac{1}{-\omega t'}\,,
\end{equation}
by using the adjoint operator (\ref{ED1}) and the new matrix-functions
\begin{equation}
{\bar\Sigma}^{(\pm)}(t,t^{\prime},\vec{x}-\vec{x}\,^{\prime}\,)=\gamma^5 {\Sigma}^{(\pm)}(t,t^{\prime},\vec{x}-\vec{x}\,^{\prime}\,)\gamma^5\,.
\end{equation} 

\section{Propagators in configuration representation}

The Dirac propagators can be calculated in any spin basis since the result is independent on the concrete spinors we use as long as the system of these spinors is complete satisfying Eq. ({\ref{Pcom}). Therefore, we will use the spin momentum-basis in the massive case and the momentum-helicity basis for massless fermions.    

\subsection{Massive case}

The principal advantage of introducing the matrix-functions $\Sigma^{(\pm)}$  is the opportunity of finding the expressions of the propagators in the configuration representation since their integrals can be solved by using Eqs. (\ref{Int}) \cite{KP}. Indeed, the integrals of Eq. (\ref{Splus1}) which are of the form
\begin{equation}
I_{\pm}(t,t',\vec{x})=\int d^3p\,e^{i\vec{p}\cdot\vec{x}}K_{\nu_{\pm}}(ipt)K_{\nu_{\pm}}(-ipt')
\end{equation} 
can be calculated in spherical coordinates in momentum space with the third axis along $\vec{x}$. Solving first the angular  integrals we remain with the radial one 
\begin{equation}
I_{\pm}(t,t',\vec{x})=\frac{4\pi}{|\vec{x}|}\int_0^{\infty}dp\, p K_{\nu_{\pm}}(ipt)K_{\nu_{\pm}}(-ipt') \sin p|\vec{x}|
\end{equation}
which has the form (\ref{Int}). Unfortunately, the condition (\ref{cond}) is not fulfilled since  in this case we have $a=it$ and $b=-it'$ such that $\Re(a+b)=0$.  Therefore we must redefine these integrals substituting $t\to t-i\epsilon$ with a small $\epsilon>0$. The new integrals
\begin{equation}\label{Int1}
I^{\epsilon}_{\pm}(t,t',\vec{x})=\frac{4\pi}{|\vec{x}|}\int_0^{\infty}dp\, p K_{\nu_{\pm}}(\epsilon p+ipt)K_{\nu_{\pm}}(-ipt') \sin p|\vec{x}|\,,
\end{equation} 
are now convergent and can be solved according to Eq. (\ref{Int}). Thus we find the definitive form of the integrals of Eq. (\ref{Splus1}) that read
\begin{equation}
I^{\epsilon}_{\pm}(t,t',\vec{x}-{\vec{x}\,}')=\frac{\pi^2}{2(tt')^{\frac{3}{2}}}\,\Gamma\left(\textstyle{\frac{3}{2}}+\nu_{\pm}\right)\Gamma\left(\textstyle{\frac{3}{2}}-\nu_{\pm}\right)F\left(\textstyle{\frac{3}{2}}-\nu_{\pm},\frac{3}{2}+\nu_{\pm};2;1+\chi_{\epsilon}\right)
\end{equation}
where 
\begin{equation}\label{chi}
\chi_{\epsilon}=\frac{(t-t'-i\epsilon)^2-(\vec{x}-{\vec{x}\,}')^2}{4tt'}\,,
\end{equation}
is related to the geodesic distance between the points $(t,\vec{x})$ and $(t', {\vec{x}\,}')$ \cite{BD}.
Thus the structure of the matrix-functions $\Sigma^{(\pm)}$ and implicitly $S^{(\pm)}$ is completely determined. For example the matrix-function (\ref{Splus1}) can be written in a closed form,
\begin{eqnarray}
&&i{\Sigma}^{(+)}_{\epsilon}(t,t^{\prime},\vec{x}-\vec{x})\nonumber\\
&&=\frac{\omega^3}{16 \pi^2}\sqrt{tt'}\left[
\pi_+ \Gamma\left(\textstyle{\frac{3}{2}}+\nu_{+}\right)\Gamma\left(\textstyle{\frac{3}{2}}-\nu_{+}\right)F\left(\textstyle{\frac{3}{2}}-\nu_{+},\frac{3}{2}+\nu_{+};2;1+\chi_{\epsilon}\right)\right.\nonumber\\
&&+\left.\pi_- \Gamma\left(\textstyle{\frac{3}{2}}+\nu_{-}\right)\Gamma\left(\textstyle{\frac{3}{2}}-\nu_{-}\right)F\left(\textstyle{\frac{3}{2}}-\nu_{-},\frac{3}{2}+\nu_{-};2;1+\chi_{\epsilon}\right)\right]\,,\label{Splusf}
\end{eqnarray}
recovering thus the result of Ref. \cite{KP} for $D=4$ and $a=-\omega t^{-1}$. Moreover, the matrix-function ${\Sigma}^{(-)}_{\epsilon}$ can be derived from Eq. (\ref{SiSi}) as 
\begin{equation}
{\Sigma}^{(-)}_{\epsilon}(t,t^{\prime},\vec{x}-{\vec{x}\,}')=-{\Sigma}^{(+)}_{\epsilon}(t^{\prime},t,\vec{x}-{\vec{x}\,}')=-{\Sigma}^{(+)}_{-\epsilon}(t,t^{\prime},\vec{x}-{\vec{x}\,}')\,,
\end{equation} 
since the expression (\ref{Splusf}) is symmetric in $t$ and $t'$ except $\chi_{\epsilon}$ for which the change $t\leftrightarrow t'$ reduces to $\epsilon \to -\epsilon$. Finally, the matrix-functions $S^{(\pm)}$ have to be calculated according to Eqs. (\ref{SSig}).

\subsection{Massless case}

 In the massless case
the  propagators can be derived directly as the limits
\begin{equation}
{S}^{(\pm)}_{0\,\epsilon}(t,t^{\prime},\vec{x}-{\vec{x}\,}^{\prime})=\lim_{\mu\to 0}\frac{1-\gamma^5}{2}\, {S}^{(\pm)}_{\epsilon}(t,t^{\prime},\vec{x}-{\vec{x}\,}^{\prime})\,\frac{1+\gamma^5}{2}\,.
\end{equation}
According to Eq. (\ref{SSig}) this can be put in the form
\begin{equation}\label{SSig1}
{S}^{(\pm)}_{0\,\epsilon}(t,t^{\prime},\vec{x}-\vec{x}\,^{\prime}\,)=\frac{1-\gamma^5}{2}\left[\frac{1}{-\omega t}D_x {\Sigma}^{(\pm)}_{0\,\epsilon}(t,t^{\prime},\vec{x}-\vec{x}\,^{\prime}\,)\right]\,,
\end{equation}
where
\begin{equation}
{\Sigma}^{(\pm)}_{0\,\epsilon}(t,t^{\prime},\vec{x}-\vec{x}\,^{\prime}\,)=\lim_{\mu\to 0}{\Sigma}^{(\pm)}_{\epsilon}(t,t^{\prime},\vec{x}-{\vec{x}\,}')\nonumber\\
=\pm i\frac{\omega^3}{16 \pi^2}\frac{\sqrt{tt'}}{\chi_{\pm\epsilon}}\,,
\end{equation}
(since $F(1,2;2;x)=(1-x)^{-1}$ and  $\Gamma(1)=\Gamma(2)=1$ \cite{NIST}). Thus we arrive at the simple final result
\begin{equation}\label{S0fin}
{S}^{(\pm)}_{0\,\epsilon}(t,t^{\prime},\vec{x}-\vec{x}\,^{\prime}\,)=\frac{1-\gamma^5}{2}\left[\pm i\,\frac{\omega^3}{16 \pi^2}\frac{1}{-\omega t}D_x \frac{\sqrt{t t'}}{\chi_{\pm\epsilon}} \right]\,,
\end{equation} 
which is different from Eq. (25) of Ref. \cite{KP}. Notice that there is not a conflict since the left-handed massless fermions are specific objects that can be defined only in $(1+3)$ dimensions having no correspondents in higher dimensions. 

An interesting exercise is to recover Eq. (\ref{S0fin}) starting directly with the spinors of the helicity basis  (\ref{n1}) and (\ref{n2}). By using the projectors defined in the Appendix B, after a few manipulation, we may put  the matrix-functions (\ref{Spm}) in the form
\begin{equation}\label{S0pm}
{S}^{(\pm)}_0(t,t^{\prime},\vec{x}-\vec{x}\,^{\prime}\,)= i \left(\frac{\omega}{2\pi}\right)^3(tt')^{\frac{3}{2}}\int d^3p\, \left(
\begin{array}{cc}
0&P_{\mp \frac{1}{2}}\\
0&0
\end{array}\right)
e^{\pm i\vec{p}\cdot(\vec{x}-{\vec{x}\,}')\mp i p(t-t')}\,,
\end{equation}
bearing in mind that here we work with the chiral representation of the Dirac matrices (with diagonal $\gamma^5$). Therefore, we may write
\begin{equation}
{S}^{(\pm)}_0(t,t^{\prime},\vec{x}-\vec{x}\,^{\prime}\,)=\frac{1-\gamma^5}{2}\left[\pm i \left(\frac{\omega}{2\pi}\right)^3\frac{1}{-\omega t}D_x (tt')^{\frac{3}{2}}I_{\pm}^0(t-t',\vec{x}-\vec{x}\,^{\prime})\right]
\end{equation}
where the integrals of the form
\begin{equation}
I_{\pm}^0(t,\vec{x})=\int \frac{d^3p}{2p}\, e^{\pm i\vec{p}\cdot(\vec{x})\mp i pt}=\frac{2\pi}{|\vec{x}|}\int_0^{\infty}dp\,e^{\mp i pt}\sin p|\vec{x}|
\end{equation}
must be replaced by the convergent ones  
\begin{equation}
I_{\pm}^0(t,\vec{x}) \to I_{\pm}^{0\,\epsilon}(t,\vec{x})=\frac{2\pi}{|\vec{x}|}\int_0^{\infty}dp\,e^{\mp i p(t\mp i\epsilon)}\sin p|\vec{x}|=- \frac{2\pi}{(t\mp i\epsilon)^2-|\vec{x}|^2}\,,
\end{equation}
which can be calculated according to Eq. (\ref{coco}) leading to the previous result (\ref{S0fin}).

\section{Concluding remarks}

In this paper we succeeded to recover the results of Ref. \cite{KP} in the massive case for $D=4$ and $a=(-\omega t)^{-1}$ but obtaining a new result for the left-handed massless fermions which are defined only in $(3+1)$ dimensions.. 

In both these cases the propagators defined by Eqs. (\ref{SR}), (\ref{SA}) and (\ref{SF}) calculated in momentum representation depend explicitly by the Heaviside functions $\theta$ such that they cannot be used for calculating transitions amplitudes in a perturbation theory as long as all the time integrals must be performed in the chronological order \cite{BDR}. 

In the Minkowski spacetime this problem is completely solved in momentum representation where the Fourier transform of the Feynman propagator includes the effects of the chronological product according to the well-know method of contour integrals \cite{BDR}. This is possible since the propagators in the flat case depend only on the difference $t-t'$ 
which plays the role of a variable of the Fourier integral. Unfortunately, this method does not work in the dS geometry where the propagators  depend explicitly on two time variables, $t$ and $t'$. We hope that a more general integral representation could solve this problem we intend to study elsewhere.

\appendix

\section{Modified Bessel functions and some integrals}

The modified Bessel functions, $K_{\nu}(z)=K_{-\nu}(z)$ \cite{GR} are real functions of complex-valued arguments related to the Hankel functions as
\begin{equation}
H^{(1/2)}_{\nu}(z)=\mp\frac{2}{\pi}e^{\mp \frac{i}{2}\pi\nu}K_{\nu}(\mp iz)\,, \quad z\in {\Bbb R}\,.
\end{equation} 
The functions  used here, $K_{\nu_{\pm}}(z)$ with
$\nu_{\pm}=\frac{1}{2}\pm i \mu$, are related among themselves through
\begin{equation}\label{H1}
[K_{\nu_{\pm}}(z)]^{*}
=K_{\nu_{\mp}}(z^*)\,,\quad \forall z \in{\Bbb C}\,,
\end{equation}
satisfy the equations
\begin{equation}\label{H2}
\left(\frac{d}{dz}+\frac{\nu_{\pm}}{z}\right)K_{\nu_{\pm}}(z)=-K_{\nu_{\mp}}(z)\,,
\end{equation}
and the identities
\begin{equation}\label{H3}
K_{\nu_{\pm}}(z)K_{\nu_{\mp}}(-z)+ K_{\nu_{\pm}}(-z)K_{\nu_{\mp}}(z)=\frac{i\pi}{ z}\,,
\end{equation}
that guarantees the correct orthonormalization properties of the fundamental spinors. For 
$\mu=0$ we have to use the simple functions \cite{NIST}
\begin{equation}\label{Km0}
K_{\frac{1}{2}}(z)=\sqrt{\frac{\pi}{2z}}e^{-z}\,.
\end{equation}

The integrals  of the form (\ref{Int1})  may be solved according to the following formula \cite{GR}
\begin{eqnarray}
&&\int_{0}^{\infty}dx\,x K_{\nu}(ax)K_{\nu}(bx) \sin cx\nonumber\\
&&~~~~~~=\frac{\pi}{8}\frac{c}{(ab)^{\frac{3}{2}}}\Gamma\left(\textstyle{\frac{3}{2}}+\nu\right)\Gamma\left(\textstyle{\frac{3}{2}}-\nu\right)F\left(\textstyle{\frac{3}{2}}-\nu,\frac{3}{2}+\nu;2;\frac{1-u}{2}\right)\label{Int}
\end{eqnarray}
where
\begin{equation}
u=\frac{a^2+b^2+c^2}{2ab}\,.
\end{equation}
Notice that these integrals are convergent for 
\begin{equation}\label{cond}
\Re(a+b)>0
\end{equation}
if $c$ is a real number. 

Another useful integral we use here is  
\begin{equation}\label{coco}
\int_{0}^{\infty}dx\,  e^{-ax}\sin bx=\frac{b}{a^2+b^2}\,, \quad \Re(a)>0\,.
\end{equation}

\section{Projectors of helicity basis}

The helicities of the massless fermions are restricted so that the spinors (\ref{n1}) and (\ref{n2}) are non-vanishing only for $\lambda=-\frac{1}{2}$ and, respectively,  $\lambda=\frac{1}{2}$. In other words, the system of spinors is incomplete which means that we must use suitable projectors when we calculate the matrix-functions (\ref{Spm}).

The normalized particle-type spinors of the helicity basis which satisfy $(\vec{\sigma}\cdot \vec{p})\,\xi_{\lambda}(\vec{p})=2\lambda\, p\, \xi_{\lambda}(\vec{p})$ have the form 
\begin{equation}
\xi_{\frac{1}{2}}(\vec{p})=\sqrt{\frac{p+p^3}{2p}}\left(
\begin{array}{c}
1\\
\frac{p^1+i p^2}{p+p^3}
\end{array}\right)\,,\quad 
\xi_{-\frac{1}{2}}(\vec{p})=\sqrt{\frac{p+p^3}{2p}}\left(
\begin{array}{c}
\frac{-p^1+i p^2}{p+p^3}\\
1
\end{array}\right)\,,
\end{equation}
while the antiparticle spinors are defined usually as $\eta_{\lambda}(\vec{p})=i\sigma_2 \xi_{\lambda}(\vec{p})^*$  \cite{BDR}. With their help one may construct the projectors  
\begin{equation}
P_{\lambda}=\xi_{\lambda}(\vec{p})\xi_{\lambda}(\vec{p})^+=\eta_{-\lambda}(\vec{p})\eta_{-\lambda}(\vec{p})^+=\frac{1}{2}+\lambda\frac{\vec{\sigma}\cdot\vec{p}}{p}\,.
\end{equation}


\begin{thebibliography}{99}


\bibitem{SW}
S. Weinberg, {\it Gravitation and Cosmology: Principles and
Applications of the General Theory of Relativity}  (Wiley, New York,
1972).

\bibitem{CR}
P. Candelas and D. J. Raine, {\em Phys. Rev. D} {\bf 12}, 965 (1975). 

\bibitem{KP}
J. F. Koksma and T. Prokopec, {\em Class. Quant. Grav.} {\bf 26}, 125003 (2009). 

\bibitem{CQED}
I. I. Cot\u aescu and C. Crucean, {\em Phys. Rev. D} {\bf 87}, 044016 (2013).

\bibitem{CD1}%momentum-helicity basis
I. I. Cot\u aescu, {\em Phys. Rev. D} {\bf 65}, 084008 (2002).

\bibitem{CD2}%energy-helicity basis
I. I. Cotaescu  and C. Crucean, {\em Int. J. Mod. Phys. A}   {\bf
23}, 3707 (2008).

\bibitem{CD3}%momentum-spin basis
I. I. Cot\u aescu, {\em Mod. Phys. Lett. A} {\bf 22}, 1613 (2011).

\bibitem{CdSquant}
I. I. Cot\u aescu, Int. J. Mod. Phys. A  {\bf 33}, 1830007 (2018).  

\bibitem{BuD}
T. S. Bunch and P. C. W. Davies, {\em Proc. R. Soc. London} {\bf 360}, 117 (1978).

\bibitem{BD}
N. D. Birrel and P. C. W. Davies,  {\em Quantum Fields in Curved
Space} (Cambridge University Press, Cambridge 1982).

\bibitem{NIST}
F. W. J. Olver, D. W. Lozier, R. F. Boisvert and C. W. Clark, {\em NIST Handbook of Mathematical Functions} (Cambridge University Press, 2010).

\bibitem{GR}
I. S. Gradshteyn and I. M. Ryzhik, {\em Table of Integrals, Series, and Products} (Academic Press, New York 2007).

\bibitem{BDR}
S. Drell and J. D. Bjorken, {\em Relativistic Quantum Fields} (Me
Graw-Hill Book Co., New York 1965).

\bibitem{Complex}
L. V. Ahlfors, {\em Complex analysis: an introduction to the theory of analytic functions of one complex variable} 
( McGraw-Hill, New York, London, 1953). 

\end{thebibliography}
\end{document}